\documentclass[iop,revtex4]{emulateapj} 

\usepackage{graphicx,subfigure,xspace}
\usepackage{float}
\usepackage{booktabs}
\usepackage{commath}
\usepackage{amsmath}
\usepackage{color}

\newcommand{\mum}{\mbox{{\usefont{U}{eur}{m}{n}{\char22}}m}\xspace}

\newcommand{\comm}[1]{}

\shorttitle{A simplified realization of a pyramid wavefront sensor}
\shortauthors{J. Lozi et al.}

\begin{document}
\title{Visible and Near Infrared Laboratory Demonstration of a Simplified Pyramid Wavefront Sensor}
\author{Julien Lozi \altaffilmark{1}, Nemanja Jovanovic\altaffilmark{2}, Olivier Guyon\altaffilmark{1,3,4,5}, Mark Chun\altaffilmark{6}, Shane Jacobson\altaffilmark{6}, Sean Goebel\altaffilmark{1,6}, and Frantz Martinache\altaffilmark{7}}
\altaffiltext{1}{Subaru Telescope, National Astronomical Observatory of Japan, National Institutes of Natural Sciences (NINS), 650 North A$\!$`oh\={o}k\={u} Place, Hilo, HI, 96720, U.S.A.}
\altaffiltext{2}{California Institute of Technology, 1200 E California Blvd, Pasadena, CA 91125, U.S.A.}
\altaffiltext{3}{Steward Observatory, University of Arizona, Tucson, AZ, 85721, U.S.A.}
\altaffiltext{4}{College of Optical Sciences, University of Arizona, Tucson, AZ 85721, U.S.A.}
\altaffiltext{5}{Astrobiology Center of NINS, 2-21-1, Osawa, Mitaka, Tokyo, 181-8588, Japan}
\altaffiltext{6}{Institute for Astronomy, University of Hawaii, 640 North A$\!$`oh\={o}k\={u} Place, Hilo, HI 96720, U.S.A.}
\altaffiltext{7}{Universit\'{e} C\^{o}te d'Azur, Observatoire de la C\^{o}te d'Azur, CNRS, Laboratoire Lagrange, 96 Boulevard de l'Observatoire, 06300 Nice, France}
\email{lozi@naoj.org}


\begin{abstract}
Wavefront sensing and control are important for enabling one of the key advantages of using large apertures, namely higher angular resolutions. Pyramid wavefront sensors are becoming commonplace in new instrument designs owing to their superior sensitivity. However, one remaining roadblock to their widespread use is the fabrication of the pyramidal optic. This complex optic is challenging to fabricate due to the pyramid tip, where four planes need to intersect in a single point. Thus far, only a handful of these have been produced due to the low yields and long lead times. To address this, we present an alternative implementation of the pyramid wavefront sensor that relies on two roof prisms instead. Such prisms are easy and inexpensive to source. We demonstrate the successful operation of the roof prism pyramid wavefront sensor on a 8-m class telescope, at visible and near infrared wavelengths ---for the first time using a SAPHIRA HgCdTe detector without modulation for a laboratory demonstration---, and elucidate how this sensor can be used more widely on wavefront control test benches and instruments.
\end{abstract}




\section{Introduction}

Implementation of new telescopes with increasing apertures is driven by the desire to increase the collecting area and hence sensitivity of an instrument, and to improve the angular resolution of the telescope. The latter can only be achieved if the telescope operates in the diffraction limited regime (i.e. in a low wavefront aberration regime), as is the case for space-based observatories. From the ground, the atmosphere corrupts the wavefront of the incoming light and if the telescope diameter is larger than the Fried parameter $r_{0}$, then the angular resolution will be limited by the seeing at that site, independently to the telescope aperture. To get around this limitation and exploit the full power of the large aperture, wavefront correction is necessary. This comes in the form of an adaptive optic (AO) system, which is used to sense and correct the wavefront in real-time, while data is being collected simultaneously. The AO system corrects for the turbulent wavefront in order to restore the flux to the core of the point-spread function (PSF) and suppress the speckle halo around the image. In this way, sources can clearly be resolved/spatially separated and fainter objects in close proximity to stars can be detected. This concept underpinned the successful Kepler follow-up observations with RoboAO \citep{law2014}, which aimed at determining if Kepler exoplanet candidates were previously unresolved binaries, background stars or other false positives. In addition, direct imaging of exoplanets from the ground-based observatories would not be possible without the speckle suppression afforded by AO systems \citep{serabyn2010,macintosh2015}.         

One of the two key components of an AO system is the wavefront sensor. There is a vast array of wavefront sensors which have a range of properties and the eager reader is directed to a comprehensive review on the topic presented by \cite{guyon2005}. In this review, the author demonstrated through simulation that a non-modulated pyramid wavefront sensor (PyWFS) \citep{raga1996} has one of the highest sensitivities of all wavefront sensors, at all spatial frequencies. To achieve its high sensitivity, the PyWFS requires a diffraction-limited PSF at the PyWFS sensing wavelength. This can be achieved either by closing the loop using the PyWFS itself, or by using an upstream wavefront control loop as a first stage of correction. If the PyWFS is modulated, which is the more typical case when used at visible wavelengths, the dynamic range of the sensor can be extended, thereby mitigating the need for an upstream AO system, but reducing at the same time the sensitivity to low-order modes. 

The benefits of such a wavefront sensor were realized early on and it has thus far been employed in several world-class AO facilities including LBTAO \citep{esposito2011}, MagAO \citep{close2013} and SCExAO \citep{jovanovic2015}. These systems have delivered many stunning images including the first images of an accreting exoplanet \citep{sallum2015} and the recent discovery of a debris disk \citep{currie2017}. For this reason PyWFSs are being considered for applications on the Giant Segmented Mirror Telescopes (GSMTs). Specifically, the Thirty Mirror Telescope (TMT) has chosen to use a PyWFS in the NFIRAOS AO instrument \citep{veran15}, the Giant Magellan Telescope (GMT) is considering it as their natural guide star wavefront sensor \citep{pinna2014} and the Extremely Large Telescope (ELT) is considering their application to the MAORY multi-conjugate AO system \citep{esposito2015} and has thus tested them in the precursor instrument MAD \citep{melnick2012}. 

Despite their potential and demonstrated successes thus far, the limitation to their widespread use has largely been associated with the difficulty in realizing the pyramidal optic at the heart of the sensor. An optic in the shape of a rectangular pyramid is typically placed in the focal plane such that the image of the source is incident on the tip of the optic and is dissected in four. These four sections of the image are reimaged to a pupil plane which is recorded on a detector. In this way, phase modulations in an upstream pupil can be encoded as intensity modulations on the detector. A key requirement of a pyramidal optic is that the apex and the vertices are as sharp as possible so that any slight motion in the image across any of those features manifests in a rapid modulation in the image intensity and hence a maximum sensitivity to a given aberration. Realizing a pyramidal optic with sharp vertices (edge defects of $<5$~\mum) is relatively easy. However, polishing an optic with four faces that meet at exactly the same location and have an apex defect which is $<5$~\mum in extent is extremely challenging. As such, only a handful of these optics have ever been created, and they were primarily made for the LBTAO and MagAO systems (J. Males, 2015, private communication). This requirement will become more extreme as PyWFS are moving towards shorter wavelengths and smaller modulations. Furthermore, the required prism angle would ideally be $<5^\circ$, and such a shallow angle is hard to manufacture. To realize this in practice, two dissimilar but steeper ($\sim30^\circ$) pyramidal optics can be cemented at their bases such that the net effect is that of a single optic with a $<5^\circ$ inclined face, compounding the challenges of fabrication. Although one advantage of this design is that two different types of glass can be used for the prisms, to make the angle of refractivity of each quadrant less chromatic.

In this work, we present a new approach to realizing the pyramidal optic. As we will show, this can be achieved through the combination of two roof prisms, which are much easier to fabricate. Section~\ref{sec:exp} describes the concept of the dual roof prism pyramid optics, and gives an overview of the experiential testbed used to validate the sensor. Section~\ref{sec:validation} presents some experimental validations of the visible PyWFS, analyzing the vertex quality and discussing about the chromaticity of this design, while Sec.~\ref{sec:NIR} presents the first demonstration of a Near Infrared (NIR) PyWFS using the dual roof prism design and a SAPHIRA detector. Some concluding statements are made in Sec.~\ref{sec:summary}. 


\section{Experimental design}\label{sec:exp}

\subsection{Concept}\label{sec:concept}

\begin{figure}
\centering 
\includegraphics[width=0.8\linewidth]{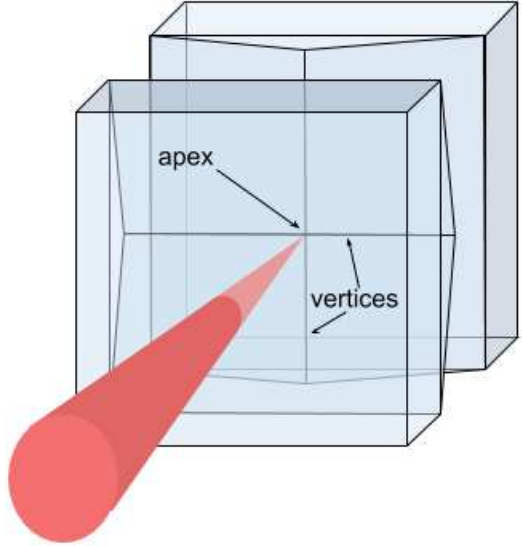}
\caption{A cartoon depicting the roof prism pyramid optic concept. The two prisms are oriented such that they meet at the vertices which are clocked at $90^{\circ}$ to one another to mimic the effect of a rectangular pyramid optic.}
\label{fig:concept}
\end{figure}

The motivation for the dual roof prism concept was to replace the classical dual pyramid loaned by MagAO for the SCExAO instrument, with minimal change in the optical setup. Two roof prisms are oriented such that the vertices are facing one another with one rotated $90^{\circ}$ with respect to the other, as seen on Fig.~\ref{fig:concept}. This creates the vertical and horizontal vertices required for the sensor.  A beam incident upon the intersection point of the two vertices will see four tilted glass faces, akin to the case of a rectangular pyramidal optic. Each face will have the effect of steering the four sub-beams inwards as required by a pyramid optic. In order to mimic the effect of a pyramidal optic, the two vertices must be in contact and the beam must be in focus at this plane. In practice however, making contact between two sharp edges on glass prisms is not advised as it could lead to their damage and subsequent light scattering. Given that a focused beam can be approximated as collimated within a length defined as the Rayleigh length, then the optics could have a gap of up to that length without affecting the performances. Using a slow beam which has a longer Rayleigh length would reduce the need for a small gap, making it easier to realize this concept without inflicting damage to the prisms or over-complicating the opto-mechanics to mount and align them. A slow beam relaxes also the constraint on the vertex quality.

Other designs have been proposed and tested as an alternative to the classical pyramid optics, such as the dual knife edge wavefront sensor \citep{ziegler2016}. This fully achromatic design uses a beamsplitter to split the light in two beams, which are focused on two pairs of knife edge mirrors that again split the light to form four pupils. This design removes any chromatic effect, at the expense of a slightly more complex design. It is not fully equivalent to a PyWFS because each pair of pupils only probes for one direction, with one pair of pupils measuring the horizontal component of the aberrations, while the second pair of pupils measures the vertical component of the aberrations. 

\subsection{Setup}\label{sec:setup}

A series of roof prisms were manufactured by IOS Optics using fused silica glass. They were $25\times25\times10$~mm in size. A roof angle (angle between the two faces of the roof) of $3.775^{\circ}$ was requested. This specification was chosen to match the net behavior of the double pyramid optic built for the MagAO and LBTAO systems \citep{tozzi2008,esposito2010} that was in use within the SCExAO instrument \citep{jovanovic2015}. The prisms were coated with an anti-reflective coating to remove any ghosting effect. The prisms were characterized and determined to meet specifications, indicating that indeed shallower optics can be fabricated in the roof prism format. It is important to mention that several manufacturers that were contacted stated they could meet our specifications, compared to the unique manufacturer of classical pyramid optics.

\begin{figure}
\centering 
\includegraphics[width=0.99\linewidth]{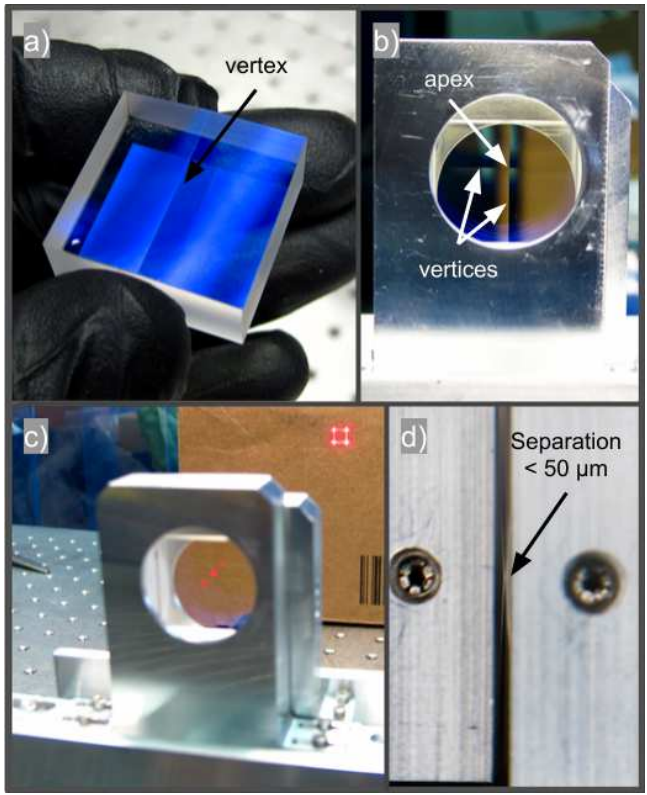}
\caption{a) A roof prism. b) An image looking through two assembled prisms. The two vertices are clearly visible. c) a collimated HeNe beam refracted by the roof prism pyramid optic into 4~spots in the far field. d) A top view of the assembled roof prism pyramid. This shows the small gap between the two prism vertices.}
\label{fig:assembly}
\end{figure}

Two roof prisms were assembled in a custom mount as shown in Fig.~\ref{fig:assembly}. The custom mounts were designed to fix the two optics with respect to one another once the inter-prism distance was set. This roof prism opto-mechanical assembly was specifically designed to replace the classical pyramid optics on loan from MagAO, without any change in the optical design. 
The beam injected into the existing PyWFS in SCExAO had a $7.2$~mm diameter, a speed of $f/40$ and a center wavelength of $850$~nm. We can approximate the Rayleigh length ($z_{R}$) for any beam to be given by 
\begin{equation}\label{rayleigh}
z_{R} \approx \frac{Cf\lambda^{2}}{2D},
\end{equation}
where $f$ is the focal length, $\lambda$ is the wavelength of the light, $D$ is the beam diameter and $C$ is a coefficient that equals $2\pi$ in the case of a Gaussian beam. From this equation, we determine that the Rayleigh length for the beam would be $\sim 680$~\mum. This would indicate that the inter-prism spacing only needs to be of this order so that the roof-prism pair would be indistinguishable to the incoming field from a regular pyramid optic. However, the inter-prism spacing was reduced to $<50$~\mum during installation to ensure proper operation without any spurious effects. The final gap between the two prisms can be seen in the bottom right panel of Fig.~\ref{fig:assembly}. 

The four pupil images are formed on the First Light OCAM2K 240x240-pixel EMCCD camera. It is used in 2x2 binned mode, to increase the maximum frame rate to 3.6~kHz. Each pupil is then imaged on a 60x60-pixel quadrant of the detector, with about 50~pixels across the diameter of the pupil. This sampling is very similar to the sampling of the deformable mirror, a Boston Micromachine 2000-actuator Deformable Mirror (DM), with 45~actuators across the pupil.

The wavefront control loop is coded using the Compute And Control for Adaptive Optics (CACAO) architecture \citep{guyon2018}. Instead of computing the slopes of the wavefront as it is commonly done for most PyWFS, the algorithm uses the whole image of the sensor, and multiplies it by a control matrix using a bank of Graphics Processing Units (GPUs) to get the DM command directly. This method has the advantage of utilizing all the light available in the sensor, and removing the necessity of need for careful registration of the DM to the detector in each pupil. A response matrix is acquired at high speed by poking all the actuators on the DM with a specific pattern, and by reconstructing the response of each DM actuator on the WFS. This matrix is inverted to form a control matrix that is used for wavefront control. When the wavefront control loop is closed, an image from the PyWFS is subtracted from a reference, and multiplied by the control matrix to form a new command map that is sent to the DM. A full description of the wavefront control architecture will be the subject of a future publication.


\section{Laboratory validation in visible}\label{sec:validation}

\subsection{Vertex quality}\label{sec:vertex}

\begin{figure}
\centering 
\includegraphics[width=0.99\linewidth]{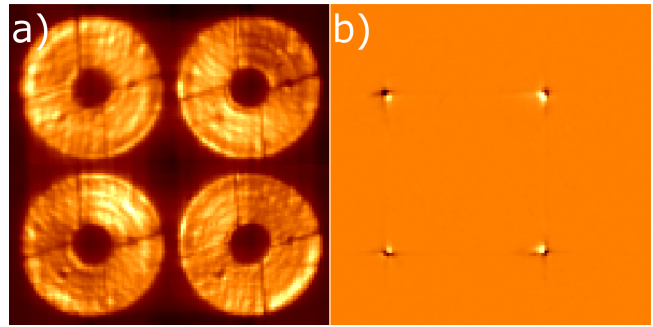}
\caption{a) Visible PyWFS image taken on-sky with the OCAM2K camera (square root scale). b) Example of response from a single DM actuator taken on-sky.}
\label{fig:vispywfs}
\end{figure}

The new roof prism pyramid successfully replaced the classical pyramid optic inside SCExAO. Only a slight change of scale of the pupil images was observed, confirming the equivalence between the divergence angles of the two designs. Figure~\ref{fig:vispywfs} (a) presents a reference image of the visible PyWFS taken on-sky, with a modulation of 75~mas ($3.4~\lambda/D$). The scale is a square root of the intensity to highlight the small amount of light scattered around the pupil by the vertices. Figure~\ref{fig:vispywfs} (b) presents the on-sky response to a single actuator poke. The response is visible on the four pupils, with only a faint cross pattern created by the vertices.

A characterization of the vertex quality of the prisms was performed by a carefully calibrated scan of the SCExAO internal source (off-sky) position using a X-Y piezo stage. The PyWFS modulation was turned off for this test. A total of 6~scans were recorded, 3 in X and 3 in Y, with different offsets on the roof prisms as outlined here: no offset (the scan goes through the apex of the pyramid), and offsets of $\pm 0.2$~arcsec, i.e. $\pm9~\lambda/D$. Finally, an average of the three scans was computed for both axes.

\begin{figure}
\centering 
\includegraphics[width=0.99\linewidth]{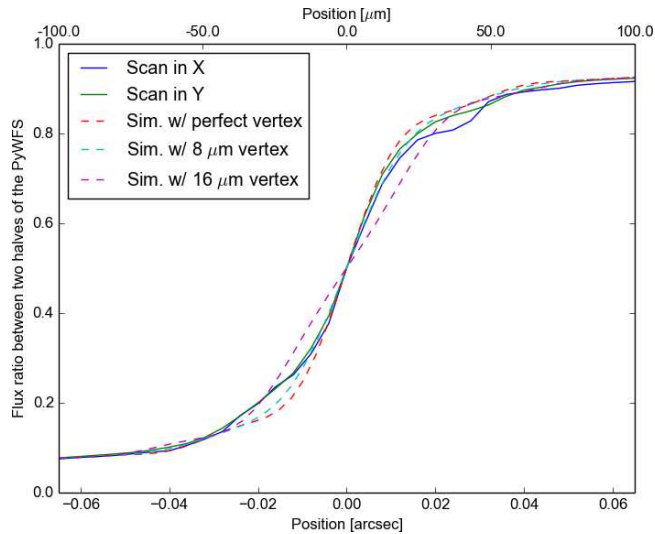}
\caption{Flux measurement in the pupils during a linear scan of the image over the vertices, in X and Y, and comparison to a simulated perfect vertex. The on-sky scale of the in the plane of the vertex is 0.65 arcsec/mm.}
\label{fig:scan}
\end{figure}

Figure~\ref{fig:scan} presents the result of this test. The metric used here is the flux ratio between a half of the PyWFS image perpendicular to the scan, and the total flux in the sensor. we get the expected functional form one would expect as we move the PSF from one side of the image to the other. The flux goes from a low level, where the measured side is not illuminated, to a high level, where the measured side gets almost all the light.

Furthermore it can be seen that the measurements are consistent with the simulations across most of the scan in both directions. Though, discrepancies are noted around the transition zones, which is not due to the quality of the vertices, but the quality of the image itself. If low-order wavefront errors are present, the image will have more flux in the first few Airy rings, or these rings will be distorted. However because the core of the image remains the same, the transition over the vertex is unchanged. As presented in Fig.~\ref{fig:scan}, a defect in the vertex would decrease the slope of the transition without affecting the tail ends of the curve much. 

From the results of Fig.~\ref{fig:scan}, we estimated that the vertex is smaller than 8~\mum, or about a quarter of the PSF core size. This is close to the specification of 5~\mum given to the manufacturer. The small vertex extent helps increase the linear range of the PyWFS in low/no modulation cases.

\subsection{Chromaticity of the roof prism pyramid}\label{sec:chromaticity}

\begin{figure}
\centering 
\includegraphics[width=0.99\linewidth]{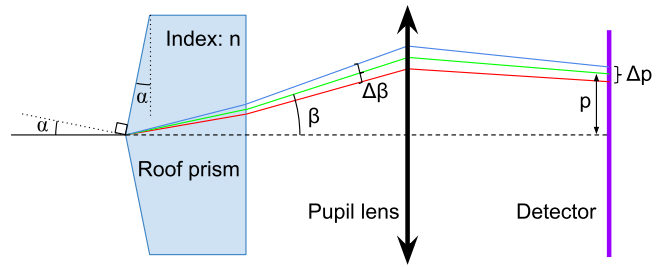}
\caption{Schematics describing the chromaticity of the roof prism pyramid. The first prism was omitted, and only the optical axis is drawn.}
\label{fig:pyrchrom}
\end{figure}

An important aspect that determines the applicability of any wavefront sensor is how it operates when a polychromatic beam is injected. Figure~\ref{fig:pyrchrom} shows the path of the rays at various wavelengths through the roof prism based wavefront sensor. It can be seen that the different wavelengths are separated on the detector. If the separation is large enough, the signal due to a wavefront aberration will be blurred in the sensor, especially if this aberration is at a large spatial frequency. A value routinely adopted as an acceptable tolerance for the chromaticity is 0.1~pixel \citep{tozzi2008,schatz2017}. A dual pyramid similar to the one used by MagAO and LBTAO can be made achromatic using two types of glass with differing dispersion, in a way similar to an achromatic doublets \citep{tozzi2008}. However, in the case of the roof prism pyramid, each prism will separate the light in only one direction. Therefore the two prisms have to be made of the same material and the same angle to keep the separation between pupils the same in X and Y, creating chromatic aberrations. 

In the case of the prisms manufactured for our system, where the angle of incidence of the light on the faces is $\alpha=3.775^\circ/2=1.8875^\circ$, the exit angle of each beam $\beta$ (see Fig.~\ref{fig:pyrchrom}) can be approximated by
\begin{equation}
    \beta\left(\lambda\right) = \alpha\left(1-n\left(\lambda\right)\right),
\end{equation}
where $n\left(\lambda\right)$ is the index of the material used for the prism.

The difference in output angle over the wavelength range used by the wavefront sensor is then simply
\begin{equation}
    \Delta\beta = \alpha\Delta n.
\end{equation}

For the SCExAO PyWFS, we use the OCAM2K camera in binned mode, i.e. 120x120~pixels. It means that the deviation angle $\beta$ in the plane of the sensor corresponds to $p \approx 30$~pixels. We can then infer the chromatic shift $\Delta p$ in that plane by scaling to this value. 

\begin{table}[]
    \centering
    \begin{tabular}{c|c|c|c|c}
        $\lambda$ range [nm] & $\beta$ [$^\circ$] & $\Delta\beta$ [$^\circ$] & $\Delta p$ (SCExAO) & $\Delta p$ (TMT) \\
        \hline
        800--900 & 0.854 & 0.0028 & 0.099~pix & 0.37~pix \\
        700--900 & 0.856 & 0.0066 & 0.23~pix & 0.86~pix \\
        600--900 & 0.857 & 0.012 & 0.41~pix & 1.5~pix 
    \end{tabular}
    \caption{Chromaticity estimations for the SCExAO roof prism pyramid, using fused silica glass.}
    \label{tab:chromaticity}
\end{table}

Table~\ref{tab:chromaticity} presents the chromaticity calculations for three wavelength ranges used routinely on SCExAO. The 800--900~nm range is the most used, while the 600--800~nm is usually used by visible science modules (VAMPIRES, RHEA and FIRST, see \cite{jovanovic2015}). In the case of fainter targets, the range of the PyWFS can be extended down to 700~nm or even 600~nm. For comparison, the table presents the same calculation done for a TMT high-contrast instrument, assuming the same sampling in the pupil for the PyWFS camera.

With SCExAO, the 800--900~nm (12\% bandwidth) range gives a chromatic shift of 0.1~pixel, within the acceptable tolerance. A wider range starting at 700~nm (25\% bandwidth) increases the shift to almost a quarter of a pixel, and the full visible range of SCExAO, i.e. 600--900~nm (40\% bandwidth), increases the chromatic shift to almost half a pixel. 

From table~\ref{tab:chromaticity}, we can see that the roof prism design may not work well for giant segmented telescopes such as TMT: the chromatic shift is already about a third of a pixel for the smallest bandwidth, and is about 1.5~pixels for a 40\% bandwidth. Nonetheless, it is possible to make this design less chromatic using two pairs of roof prisms instead of one, with different type of glass, as suggested in \cite{chen2017} for an infrared PyWFS for the MMT Observatory.

Since the chromaticity is the main issue for this design, we analyzed response matrices taken with the different wavelength bands from Tab.~\ref{tab:chromaticity}. We acquired response matrices with the different dichroics while the PyWFS was running at 2~kHz, with a modulation of 125 mas ($\sim 6\lambda/D$), giving us similar results as in Fig.~\ref{fig:vispywfs}~(b). For all three cases, the response matrices acquired had high signal-to-noise ratios. We multiplied the resulting matrices by simulated DM maps comprised of a one-dimensional sine wave with an increasing spatial frequency. Finally, we calculated the standard deviation in the resulting response as a metric of the sensitivity. 

\begin{figure}
\centering 
\includegraphics[width=0.99\linewidth]{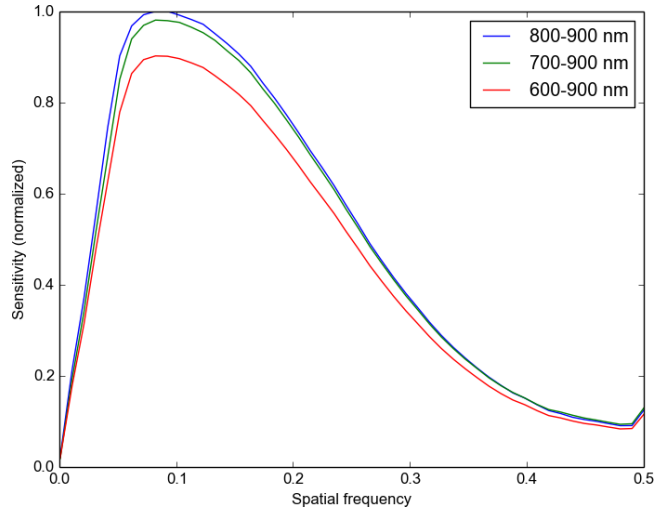}
\caption{Sensitivity of the modulated visible PyWFS for different spatial frequencies and different wavelength bands. A spatial frequency of 0.5 corresponds to the highest spatial frequency achievable with the 50x50 DM, with 25 cycles over the DM.}
\label{fig:sens}
\end{figure}
 
The result of relative sensitivity versus spatial frequency applied on the DM is presented in Fig.~\ref{fig:sens}. In this figure, we notice the same sensitivity profile for the three wavelength bands, with a linear increase for spatial frequencies between 0 and about 0.06 ($\sim 3$~cycles per aperture), then a decrease up to the highest spatial frequency. The first increase corresponds to the decrease in sensitivity from the modulation of the PyWFS, with a linear behavior described in \cite{verinaud2004}, while the second part corresponds to the natural decrease in sensitivity due to the lower sampling over the sine waves in the PyWFS. It is important to note that the sensitivity is only slightly affected by the increased bandwidth: increasing the bandwidth from 12\% to 25\% only reduced the sensitivity by a few percent, while increasing it to 40\% reduced the sensitivity by about 10\%. This result shows that despite the chromatic effect described above, the PyWFS can easily close the loop even for a wide bandwidth. 

The visible PyWFS performed successfully on-sky during multiple engineering and science observations, in various conditions and on targets brighter than $m_R \approx 9$, where Strehl ratios higher than 80\% were routinely achieved \citep{currie2017,currie2018,kuhn2018,goebel2018b}.  The loop was closed with a 40\% spectral bandwidth on targets brighter than $m_R \approx 12$, and offered a significant improvement of the Strehl ratio. The details of the on-sky performance of the visible PyWFS will be the subject of a separate publication.


\section{First demonstration of the non-modulated NIR PyWFS using a SAPHIRA detector}\label{sec:NIR}

Using PyWFS systems in NIR is a necessary step to image exoplanets around redder stars such as M-type stars, that are too faint in the visible. Longer wavelengths mean that aberrations are easier to measure, and the modulation can be reduced or even stopped.  Limitations in detector technologies restrained the development of NIR PyWFS until recently. Although the first on-sky demonstration of such a system was performed using a HAWAII I detector with PYRAMIR, mounted on the ALFA adaptive optics at the 3.5~m telescope of the Calar Alto Observatory \citep{peter2010}, the correction obtained ---mainly in K-band--- was not enough to justify its cost on other systems. It is only with the recent developments of the fast NIR HgCdTe SAPHIRA detector by Leonardo that efficient NIR PyWFS are now possible.

The roof-prism pyramid system was combined for the first time at the end of 2015 with a SAPHIRA detector, deployed by the University of Hawaii Institute for Astronomy \citep{goebel2018a}, for a quick demonstration of the power of combining the two technologies. This demonstration was intended to be a proof of concept for the Keck Planet Imager and Characterizer (KPIC) instrument \citep{mawet2016}. An optical system was added in front of the camera, to replicate the visible PyWFS, but this time at 1600~nm instead of 850~nm (bandwidth of 20\%). The same algorithms as those used for the visible PyWFS (see Sec.~\ref{sec:setup}) ---response matrix acquisition, as well as control matrix computation and real-time control using the full image instead of intensity differences--- were used for this test.

\begin{figure}
\centering 
\includegraphics[width=0.99\linewidth]{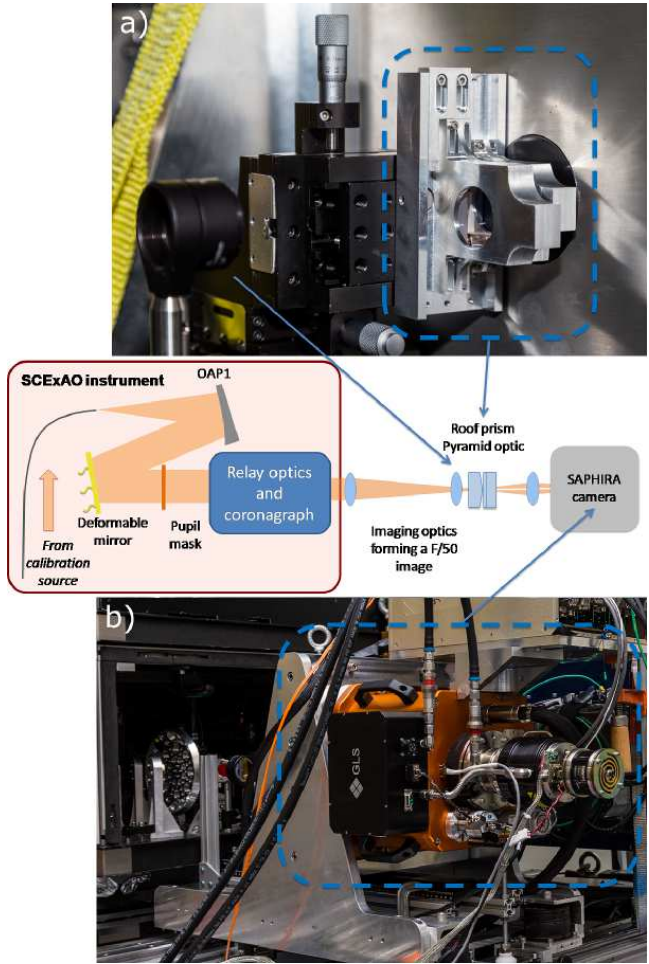}
\caption{Schematic of the NIR PyWFS demonstration using a) a second set of roof prism pyramid optics installed in front of b) the SAPHIRA camera deployed by the University of Hawaii Institute for Astronomy.}
\label{fig:saphira}
\end{figure}

Figure~\ref{fig:saphira} presents a schematic of the setup for the NIR PyWFS test: a pair of lenses focused the collimated light coming from one of the outputs behind the SCExAO instrument on the apex of the roof-prism pyramid optics, with a speed of $f/50$. The roof-prism pyramid optics were installed on an X-Y mount for precise alignment of the beam on the apex (Fig.~\ref{fig:saphira} (a)). Finally, a pupil lens was used to re-image the four pupils on to the SAPHIRA camera (Fig.~\ref{fig:saphira} (b)).

The SAPHIRA camera was read out using a LEACH controller \citep{leach2000,goebel2016}, limiting the loop update rate to 120~Hz. The test was performed using SCExAO's internal broadband source (super continuum laser), as well as on-sky. We will present only the laboratory results here, since the frame rate of the camera was too slow to show any improvement on-sky. In combination with the internal source, turbulence was added on the deformable mirror to simulate atmospheric errors (see \cite{jovanovic2015} for a full description of the turbulence simulator). This turbulence has a Kolmogorov profile, a wind speed of 5~m/s, with low-order modes amplified. The wavefront error amplitude was set to about 400~nm RMS, although the real value is unknown due to non-linear effects of the DM.

Because the sensing wavelength was about twice longer than that of the visible PyWFS, the range of linearity also increased. Therefore the test was performed without modulation, to exploit the full sensitivity of the NIR PyWFS for low spatial frequencies. A tip/tilt alignment loop was also implemented to correct for any slow drifts of the image on the apex of the roof-prism pyramid.

\begin{figure}
\centering 
\includegraphics[width=0.99\linewidth]{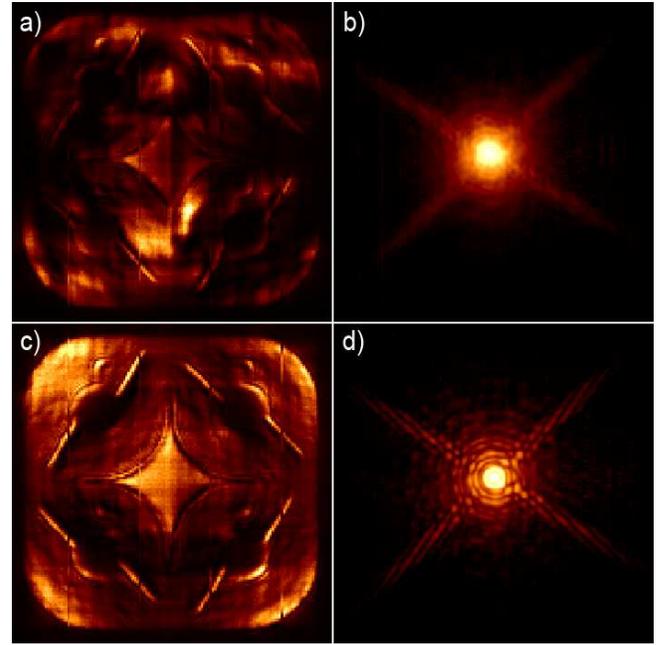}
\caption{a) PyWFS pupil image with the loop open with simulated turbulence applied. b) Loop open, focal plane image. c) PyWFS pupil image with the loop closed at 110~Hz. d) Loop closed, focal plane image. The pupil images were acquired using the full H-band, while the focal plane images were taken in the 1.45--1.7~\mum wavelength band.}
\label{fig:NIRloop}
\end{figure}

Figure~\ref{fig:NIRloop} presents results of the laboratory test, in open-loop (top) and in closed-loop (bottom). We present instantaneous images (8.3~ms) in the NIR PyWFS (left), as well as an average of 1000~frames of the focal plane Short-Wave IR (SWIR, 1.45--1.7~\mum) camera simulating a $\sim 6$~s exposure time (right). The PyWFS pupil images are automatically scaled using their respective minimum and maximum values, while the focal plane images have the same scale. Figure~\ref{fig:NIRloop} (d) was deliberately saturated to look at the speckle field.

This figure shows that closing the loop greatly improves the uniformity of the illumination of the pupils on the PyWFS image, as well as the speckle field sharpness in the focal plane image. It can be noted that the pupil images are almost overlapping, due to a slight error in the optical design of the proof of concept system. Some light can be seen between the pupils, which is characteristic of a non-modulated PyWFS, but it is not brighter than the rest of the image. Once again, this indicates a good vertex quality for both prisms.

\begin{figure}
\centering 
\includegraphics[width=0.99\linewidth]{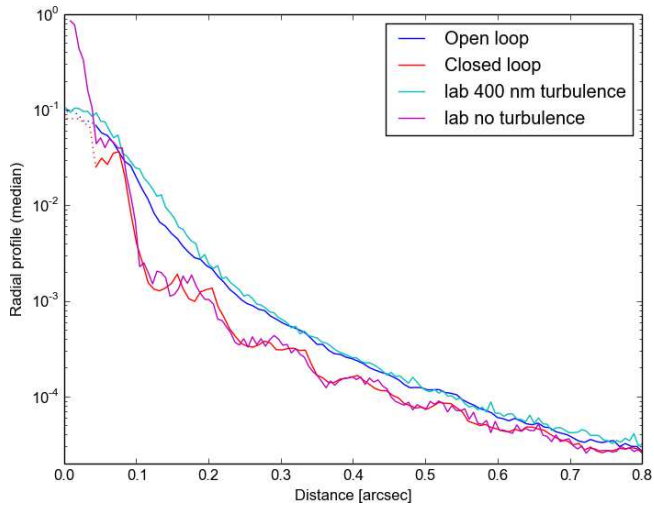}
\caption{PSF radial profiles in open- and closed-loop, compared to recent laboratory measurements. All data were taken in the 1.45--1.7~\mum wavelength band.}
\label{fig:profile}
\end{figure}

From the focal plane images, we compared the median radial profile in open- and closed-loop to recent laboratory PSFs, with and without the $\sim400$~nm of turbulence added on the deformable mirror. The comparison is presented in Fig.~\ref{fig:profile}. Unfortunately, the data taken during the NIR pyramid test were saturated (dotted lines), so the Strehl ratio in closed-loop can only be estimated from the comparison with the unsaturated laboratory images. The open-loop profile matches the profile of the laboratory image acquired by adding the same turbulence on the DM. So in open-loop, the Strehl ratio is estimated to be about 10\%. In closed-loop, the PSF profile matches the laboratory PSF when no turbulence is applied. In closed-loop the Strehl ratio is then about 90\%, demonstrating that closing the loop with the NIR PyWFS removes almost all of the turbulence added on the DM.

The SAPHIRA detector, combined with the new Pizzabox electronics ---replacing the LEACH controller---, can now acquire full frame images at 380~Hz and $128\times128$-pixel sub-window images at 1.68~kHz, making it fast enough for on-sky implementation \citep{goebel2018a}.  A SAPHIRA detector is also used in the FirstLight imaging C-RED ONE camera, that can read the full frame image at 3.5~kHz, and a $128\times128$ image at about 15~kHz. This camera is ideal for a NIR PyWFS, and will potentially be used for an upgrade of SCExAO. This brief demonstration, undertaken back in 2015, seeded the development of a dedicated NIR PyWFS for KPIC, which has recently begun commissioning \citep{bond2018}.


\section{Summary}\label{sec:summary}
In this paper, we demonstrate the application of a pair of roof prisms to simulate a pyramidal optic for pyramid wavefront sensing. This approach simplifies the fabrication process of the optics, reducing the cost and lead time greatly, which will enable numerous new implementations on high-contrast test benches and instruments. We validated that the roof prisms were made to specifications from a standard manufacturer, and once installed, behaved as a single pyramidal optic. The PyWFS was tested within the SCExAO instrument and it was shown that we could successfully close the loop in a laboratory setting, as well as on-sky. Despite increased chromatic effects from classical pyramid optics, we showed that the double roof prism design, even using visible wavelengths, didn't impact significantly the closed-loop performance on a 8-m class telescope. Therefore, for 30-m class telescopes such as TMT, the chromaticity shouldn't be a problem for small bandwidths up to 12\%, while it is possible to use a pair of double roof prisms to reduce the chromaticity at acceptable levels for larger bandwidths. Finally, we demonstrated for the first time the closed-loop operation of a non-modulated NIR PyWFS using the same roof-prism pyramid and a fast NIR SAPHIRA detector. This configuration is now used in the new KPIC instrument at the Keck Telescope. This new approach to the pyramid optics will simplify the implementation of pyramid wavefront sensors on other testbeds and instruments, and make them accessible to GSMTs and other telescopes.   

\acknowledgments
The development of SCExAO was supported by the Japan Society for the Promotion of Science (Grant-in-Aid for Research \#23340051, \#26220704 \& \#23103002), the Astrobiology Center of the National Institutes of Natural Sciences, Japan, the Mt Cuba Foundation and the director's contingency fund at Subaru Telescope. F. Martinache's work is supported by the ERC award CoG - 683029. The authors wish to recognize and acknowledge the very significant cultural role and reverence that the summit of Maunakea has always had within the indigenous Hawaiian community.  We are most fortunate to have the opportunity to conduct observations from this mountain.

\textit{Facilities: Subaru Telescope, NAOJ.}


\begin{thebibliography}{}
\bibitem[Bond et al. (2018)]{bond2018} Bond, C. Z., Wizinowich, P., Chun, M., et al.\ 2018, \procspie, 10703, 107031Z
\bibitem[Chen et al. (2017)]{chen2017} Chen, S., Sivanandam, S., Liu, S., et al.\ 2017, \procspie, 10401, 104011H
\bibitem[Close et al. (2013)]{close2013} Close, L.~M., Males, J.~R., Morzinski, K., et al.\ 2013, \apj, 774, 94 
\bibitem[Currie et al. (2017)]{currie2017} Currie, T., Guyon, O., Tamura, M., et al.\ 2017, \apjl, 836, 1
\bibitem[Currie et al. (2018)]{currie2018} Currie, T., Brandt, T., Uyama, T., et al.\ 2018, \aj, 156, 6
\bibitem[Esposito et al. (2010)]{esposito2010} Esposito, S., Riccardi, A., Fini, L., et al.\ 2010, \procspie, 7736, 773609 
\bibitem[Esposito et al. (2011)]{esposito2011} Esposito, S., Riccardi, A., Pinna, E., et al.\ 2011, \procspie, 8149, 814902 
\bibitem[Esposito et al. (2015)]{esposito2015} Esposito, S., Agapito, G., Antichi, J., et al.\ 2015, \memsai, 86, 446
\bibitem[Goebel et al. (2016)]{goebel2016} Goebel S. B., Guyon, O., Hall, D. N. B., et al.\ 2016, \procspie, 9909, 990919
\bibitem[Goebel et al. (2018a)]{goebel2018a} Goebel, S. B., Hall, D. N. B., Guyon, O., et al.\ 2018, J. Astron. Telesc. Instrum. Syst., 4(2), 026001
\bibitem[Goebel et al. (2018b)]{goebel2018b} Goebel, S. B.,  Currie, T., Guyon, O., et al.\ 2018, \aj, 156, 6
\bibitem[Guyon (2005)]{guyon2005} Guyon, O.\ 2005, \apj, 629, 592
\bibitem[Guyon et al. (2018)]{guyon2018} Guyon, O., Sevin, A., Gratadour, D., et al.\ 2018, \procspie, 10703, 107031E
\bibitem[Jovanovic et al. (2015)]{jovanovic2015} Jovanovic, N., Martinache, F., Guyon, O., et al.\ 2015, \pasp, 127, 890 
\bibitem[K\"{u}hn et al. (2018)]{kuhn2018} K\"{u}hn, J., Serabyn, E., Lozi, J., et al.\ 2018, \pasp, 130, 985 
\bibitem[Law et al. (2014)]{law2014} Law, N.~M., Morton, T., Baranec, C., et al.\ 2014, \apj, 791, 35 
\bibitem[Leach \& Low (2000)]{leach2000} Leach, R.~W., \& Low, F.~J.\ 2000, \procspie, 4008, 337
\bibitem[Macintosh et al. (2015)]{macintosh2015} Macintosh, B., Graham, J.~R., Barman, T., et al.\ 2015, Science, 350, 64
\bibitem[Mawet et al. (2016)]{mawet2016} Mawet, D., Wizinowich, P., Dekany, R., et al.\ 2016, \procspie, 9909, 99090D
\bibitem[Melnick et al. (2012)]{melnick2012} Melnick, J., Marchetti, E., \& Amico, P.\ 2012, \procspie, 8447, 84470M
\bibitem[Peter et al. (2010)]{peter2010} Peter, D. Feldt, M., Henning, T., et al.\ 2010, \pasp, 122, 887
\bibitem[Pinna et al. (2014)]{pinna2014} Pinna, E., Agapito, G., Quir{\'o}s-Pacheco, F., et al.\ 2014, \procspie, 9148, 91482M
\bibitem[Ragazzoni (1996)]{raga1996} Ragazzoni, R. \ 1996, J. of Mod. Opt., 43, 289
\bibitem[Serabyn et al. (2010)]{serabyn2010} Serabyn, E., Mawet, D., \& Burruss, R.\ 2010, \nat, 464, 1018
\bibitem[Sallum et al.(2015)]{sallum2015} Sallum, S., Follette, K.~B., Eisner, J.~A., et al.\ 2015, \nat, 527, 342  
\bibitem[Schatz et al. (2017)]{schatz2017}Schatz, L., Durney, O., Males, J. R., et al.,\ 2017, Proc. of AO4ELT5
\bibitem[Tozzi et al. (2008)]{tozzi2008} Tozzi, A., Stefanini, P., Pinna, E., Esposito, S.\ 2008, \procspie, 7015, 701558
\bibitem[V\'{e}ran et al. (2015)]{veran15} V\'{e}ran, J.-P., Esposito, S., Spano, P., Herriot, G., \& Andersen, D.,\ 2015, Proc. of AO4ELT4
\bibitem[V\'{e}rinaud (2004)]{verinaud2004} V\'{e}rinaud, C.\ 2004, \oc, 233, 1-3
\bibitem[Ziegler et al. (2016)]{ziegler2016} Ziegler, C., Law, N. M., Tokovinin, A.\ 2016, \procspie, 9909, 99093Z


\end{thebibliography}
\end{document}